\documentclass[twocolumn,showpacs,preprintnumbers,amsmath,amssymb,superscriptadd
ress]{revtex4}

\usepackage{graphicx}
\usepackage{dcolumn}
\usepackage{bm}

\def\bea{\begin{eqnarray}}
\def\eea{\end{eqnarray}}

\begin{document}

\preprint{Version 2.7}

\title{Minijet deformation and charge-independent angular correlations on momentum subspace $(\eta,\phi)$ in Au-Au collisions at $\sqrt{s_{NN}}$ = 130 GeV}

\affiliation{Argonne National Laboratory, Argonne, Illinois 60439}
\affiliation{University of Bern, 3012 Bern, Switzerland}
\affiliation{University of Birmingham, Birmingham, United Kingdom}
\affiliation{Brookhaven National Laboratory, Upton, New York 11973}
\affiliation{California Institute of Technology, Pasadena, California 91125}
\affiliation{University of California, Berkeley, California 94720}
\affiliation{University of California, Davis, California 95616}
\affiliation{University of California, Los Angeles, California 90095}
\affiliation{Carnegie Mellon University, Pittsburgh, Pennsylvania 15213}
\affiliation{Creighton University, Omaha, Nebraska 68178}
\affiliation{Nuclear Physics Institute AS CR, 250 68 \v{R}e\v{z}/Prague, Czech Republic}
\affiliation{Laboratory for High Energy (JINR), Dubna, Russia}
\affiliation{Particle Physics Laboratory (JINR), Dubna, Russia}
\affiliation{University of Frankfurt, Frankfurt, Germany}
\affiliation{Institute of Physics, Bhubaneswar 751005, India}
\affiliation{Indian Institute of Technology, Mumbai, India}
\affiliation{Indiana University, Bloomington, Indiana 47408}
\affiliation{Institut de Recherches Subatomiques, Strasbourg, France}
\affiliation{University of Jammu, Jammu 180001, India}
\affiliation{Kent State University, Kent, Ohio 44242}
\affiliation{Lawrence Berkeley National Laboratory, Berkeley, California 94720}
\affiliation{Massachusetts Institute of Technology, Cambridge, MA 02139-4307}
\affiliation{Max-Planck-Institut f\"ur Physik, Munich, Germany}
\affiliation{Michigan State University, East Lansing, Michigan 48824}
\affiliation{Moscow Engineering Physics Institute, Moscow Russia}
\affiliation{City College of New York, New York City, New York 10031}
\affiliation{NIKHEF and Utrecht University, Amsterdam, The Netherlands}
\affiliation{Ohio State University, Columbus, Ohio 43210}
\affiliation{Panjab University, Chandigarh 160014, India}
\affiliation{Pennsylvania State University, University Park, Pennsylvania 16802}
\affiliation{Institute of High Energy Physics, Protvino, Russia}
\affiliation{Purdue University, West Lafayette, Indiana 47907}
\affiliation{Pusan National University, Pusan, Republic of Korea}
\affiliation{University of Rajasthan, Jaipur 302004, India}
\affiliation{Rice University, Houston, Texas 77251}
\affiliation{Universidade de Sao Paulo, Sao Paulo, Brazil}
\affiliation{University of Science \& Technology of China, Hefei 230026, China}
\affiliation{Shanghai Institute of Applied Physics, Shanghai 201800, China}
\affiliation{SUBATECH, Nantes, France}
\affiliation{Texas A\&M University, College Station, Texas 77843}
\affiliation{University of Texas, Austin, Texas 78712}
\affiliation{Tsinghua University, Beijing 100084, China}
\affiliation{Valparaiso University, Valparaiso, Indiana 46383}
\affiliation{Variable Energy Cyclotron Centre, Kolkata 700064, India}
\affiliation{Warsaw University of Technology, Warsaw, Poland}
\affiliation{University of Washington, Seattle, Washington 98195}
\affiliation{Wayne State University, Detroit, Michigan 48201}
\affiliation{Institute of Particle Physics, CCNU (HZNU), Wuhan 430079, China}
\affiliation{Yale University, New Haven, Connecticut 06520}
\affiliation{University of Zagreb, Zagreb, HR-10002, Croatia}

\author{J.~Adams}\affiliation{University of Birmingham, Birmingham, United Kingdom}
\author{M.M.~Aggarwal}\affiliation{Panjab University, Chandigarh 160014, India}
\author{Z.~Ahammed}\affiliation{Variable Energy Cyclotron Centre, Kolkata 700064, India}
\author{J.~Amonett}\affiliation{Kent State University, Kent, Ohio 44242}
\author{B.D.~Anderson}\affiliation{Kent State University, Kent, Ohio 44242}
\author{D.~Arkhipkin}\affiliation{Particle Physics Laboratory (JINR), Dubna, Russia}
\author{G.S.~Averichev}\affiliation{Laboratory for High Energy (JINR), Dubna, Russia}
\author{S.K.~Badyal}\affiliation{University of Jammu, Jammu 180001, India}
\author{Y.~Bai}\affiliation{NIKHEF and Utrecht University, Amsterdam, The Netherlands}
\author{J.~Balewski}\affiliation{Indiana University, Bloomington, Indiana 47408}
% ...

\collaboration{STAR Collaboration}\noaffiliation

\date{\today}

\begin{abstract}
First measurements of charge-independent correlations on angular difference variables $\eta_1 - \eta_2$ (pseudorapidity) and $\phi_1 - \phi_2$ (azimuth) are presented for primary charged hadrons with transverse momentum $0.15 \leq p_t \leq 2$~GeV/$c$ and $|\eta| \leq 1.3$ from Au-Au collisions at $\sqrt{s_{NN}} = 130$~GeV. Strong charge-independent angular correlations are observed associated with jet-like structures and elliptic flow. The width of the jet-like peak on $\eta_1 - \eta_2$ increases by a factor 2.3 from peripheral to central collisions, suggesting strong coupling of semi-hard scattered partons to a longitudinally-expanding medium. New methods of jet analysis introduced here provide evidence for nonperturbative QCD medium effects in heavy ion collisions.
\end{abstract}


\maketitle

\section{Introduction}

Correlations and fluctuations can provide essential information on the nature of the medium produced in ultrarelativistic heavy ion collisions~\cite{stock,poly,dcc}. {\em In-medium modification} of parton scattering and fragmentation of energetic partons and the bulk medium can be observed {\em via} large-momentum-scale correlations. Charge-independent angular correlations result from initial-state multiple scattering (Cronin effect~\cite{iss}, hard parton scattering~\cite{jetquench}) with subsequent in-medium parton dissipation~\cite{newref} and elliptic flow. Medium modification of minimum-bias, semi-hard parton scattering and fragmentation ({\em i.e., minijets}) is the subject of this paper. 

Previous studies of parton-medium interactions have included angular correlations of high-$p_t$ particles based on a {\em leading-particle} technique ({\em e.g.,} leading-particle $p_t > 4$ GeV/c, associated particle $p_t < 4$ GeV/c) in which the away-side jet structure was observed to be strongly reduced in central Au-Au collisions~\cite{backjet}. Theoretical descriptions of parton energy loss and medium-modified fragmentation include perturbative quantum chromodynamics (pQCD) based jet-quenching models~\cite{jetquench,salgado} and parton recombination models~\cite{reco}.

In this Letter we report the first measurements in heavy ion collisions of charge-independent {\em joint autocorrelations}~\cite{auto} on angular difference variables $\phi_{\Delta} \equiv \phi_1 - \phi_2$ (azimuth) {\em and}\, $\eta_{\Delta} \equiv \eta_1 - \eta_2$ (pseudorapidity) for charged particles with $0.15 \leq p_t \leq 2$ GeV/c. This analysis is based on $\sqrt{s_{NN}} = 130$~GeV Au-Au collisions observed with the STAR detector~\cite{star} at the Relativistic Heavy Ion Collider (RHIC). 
The measurements involve no {\em a priori} jet model (specifically, no trigger particle). 
The autocorrelation technique, combined with the large angular acceptance of the STAR detector, enables statistically weak correlation structures, which are individually undetectable but occur multiple times in each event, to be measured in the aggregate with good statistical accuracy. 
These novel {\em low}-$p_t$ measurements reveal jet-like correlations which suggest that in central Au-Au collisions strong non-perturbative coupling of partons to a {\em longitudinally-expanding medium}~\cite{schukraft} produces dramatic changes in the angular distribution of parton fragments (hadrons) not anticipated by pQCD-based theory~\cite{jetquench,rqmd}. 

\section{Analysis Method}

Our goal is to access the complete {\em charge-independent} (CI - all charged particles) structure of two-particle density $\rho(\vec{p_1},\vec{p_2})$ {\em without imposing a correlation model}. The full two-particle momentum space is projected onto angular subspace $(\eta_1,\eta_2,\phi_1,\phi_2)$, integrating over a specific transverse momentum interval. Correlation structure on {transverse momentum} with specific angular constraints is considered in a separate analysis~\cite{mtxmt}. We further project the 4D (four dimensional) subspace onto a 2D subspace of angular {\em difference variables} to form a {\em joint autocorrelation} which is observed to retain essentially all angular correlation information~\cite{cdpaper}. The autocorrelations obtained in this study thus access the {\em complete CI angular structure} of two-particle density $\rho(\vec{p_1},\vec{p_2})$.

Differential correlation analysis is achieved by comparing an object distribution $\rho_{sib}$ of particle pairs taken from single events (sibling pairs) with a reference distribution $\rho_{mix}$ where each particle in the pair is taken from different but similar events (mixed pairs). The corresponding correlation function and pair-number density ratio are defined by
\bea
C(\vec{p}_1,\vec{p}_2) & = &
\rho_{sib}(\vec{p}_1,\vec{p}_2) - \rho_{mix}(\vec{p}_1,\vec{p}_2) \nonumber \\
r(\vec{p}_1,\vec{p}_2) & \equiv & \rho_{sib}(\vec{p}_1,\vec{p}_2)/
\rho_{mix}(\vec{p}_1,\vec{p}_2).
\label{Eq1}
\eea
Pair densities $\rho(\vec{p}_1,\vec{p}_2)$ are first projected onto subspaces $(\eta_1,\eta_2)$, $(\phi_1,\phi_2)$ as histograms $n_{ab} \simeq  \epsilon_x \, \epsilon_y \,\rho(x_a,y_b)$, where $ab$ are 2D bin indices and $\epsilon_x, \epsilon_y$ are bin widths on $x,y \in \{\eta, \phi, \eta_\Delta, \phi_\Delta\}$. Sibling- and mixed-pair histograms are separately normalized to the total number of detected pairs in each event class: $\hat n_{ab} = n_{ab} / \sum_{ab} n_{ab}$. Normalized pair-number ratios $\hat{r}_{ab} = \hat{n}_{ab,sib}/\hat{n}_{ab,mix}$ are the basis for this analysis. Ratios are formed from subsets of events with similar centrality (multiplicities differ by $\leq 100$, except $\leq 50$ for most-central) and primary-vertex location (within 7.5~cm along the beam axis) and combined as weighted (by sibling pair number) averages within each centrality class.
If correlation structure is invariant on sum variables $\eta_1 + \eta_2$ and $\phi_1 + \phi_2$ (stationarity), as observed previously in heavy ion collisions~\cite{cdpaper}, histograms $\hat r_{ab}$ may be projected {\em by averaging} along parallel diagonals to form 2D {joint}\, autocorrelations~\cite{auto,inverse} on difference variables $\eta_\Delta , \phi_\Delta$. Autocorrelation details are described in~\cite{inverse,mitmeths}. The autocorrelations in this study should not be confused with {\em conditional distributions} obtained from leading-particle jet analyses. 2D joint autocorrelations compactly represent all angular correlations on 4D subspace $(\eta_1,\eta_2,\phi_1,\phi_2)$ {\em without information loss} (distortion).

\begin{figure}[t]
\includegraphics[keepaspectratio,width=3.3in]{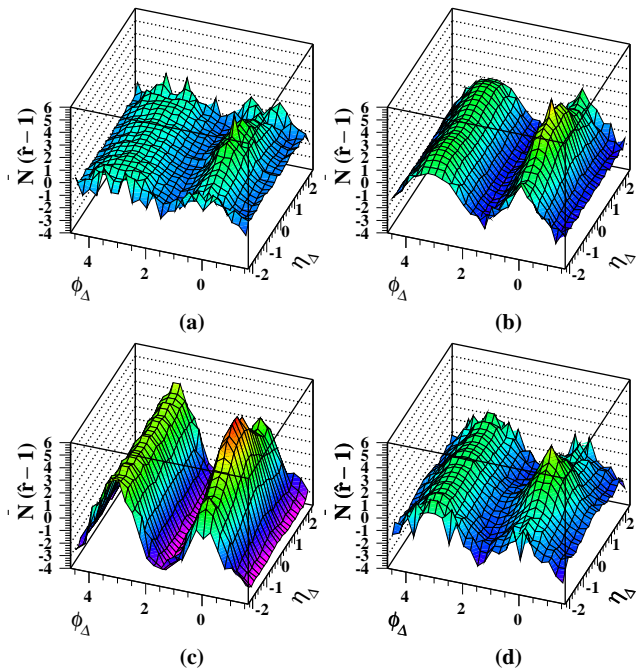}
\caption{\label{Figure1}
Perspective views of two-particle CI joint autocorrelations $\bar N (\hat r-1)$ on  $(\eta_{\Delta},\phi_{\Delta})$ for central (a) to peripheral (d) collisions.}
\end{figure}

\section{Data}
\label{SecData}

Data for this analysis were obtained with the STAR detector~\cite{star}
using a 0.25~T uniform magnetic field parallel to the beam axis. Event triggering and charged-particle measurements with the time projection chamber (TPC) are described in \cite{star}. Track definitions, tracking efficiencies, quality cuts and primary-particle definition are described in~\cite{meanptprl,spectra}.  Tracks were accepted in $|\eta| \leq 1.3$, $0.15 \leq p_t \leq 2$~GeV/$c$ and full azimuth. Particle identification was not implemented. Corrections were made to $\hat{r}$ for two-track inefficiencies due to track merging and intersecting trajectories reconstructed as $>2$ particles (splitting)~\cite{trackcuts}. Small-scale momentum correlations due to HBT and Coulomb effects~\cite{starhbt} were suppressed by eliminating sibling and mixed track pairs with $|\eta_{\Delta}| < 0.3$, $|\phi_{\Delta}| < \pi/6$, $|p_{t1} - p_{t2}| < 0.15$~GeV/$c$, if $p_t < 0.8$~GeV/$c$ for either particle. These pair cuts have negligible effect on the correlations studied here. Four centrality classes for 300k events labeled (a) - (d) for central to peripheral were defined by cuts on TPC track multiplicity $N$ within the acceptance relative to minimum-bias event multiplicity frequency distribution end-point $N_0$~\cite{endpoint}, which corresponds to the maximum participant number~\cite{meanptprl,nu}. The four centrality classes used here were defined by (d) $0.03 < N/N_0 \leq 0.21$, (c) $0.21 < N/N_0 \leq 0.56$, (b) $0.56 < N/N_0 \leq 0.79$ and (a) $N/N_0 > 0.79$, corresponding respectively to fraction of total cross section ranges 40-70\%, 17-40\%, 5-17\% and 0-5\%.


\begin{figure}[t]
\includegraphics[keepaspectratio,width=1.65in]{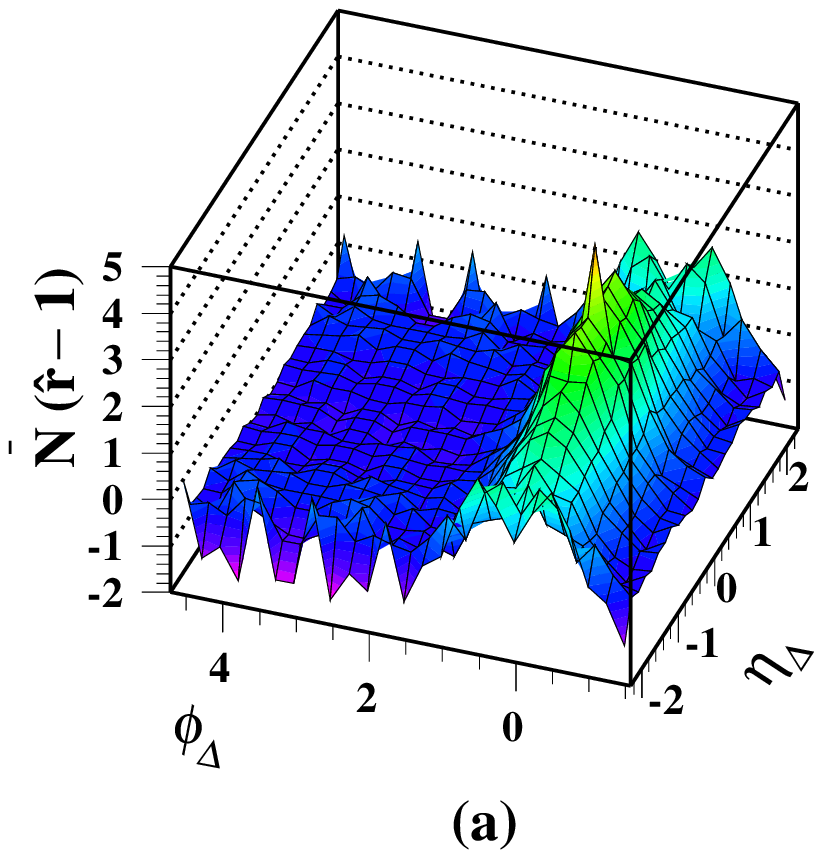}
\includegraphics[keepaspectratio,width=1.65in]{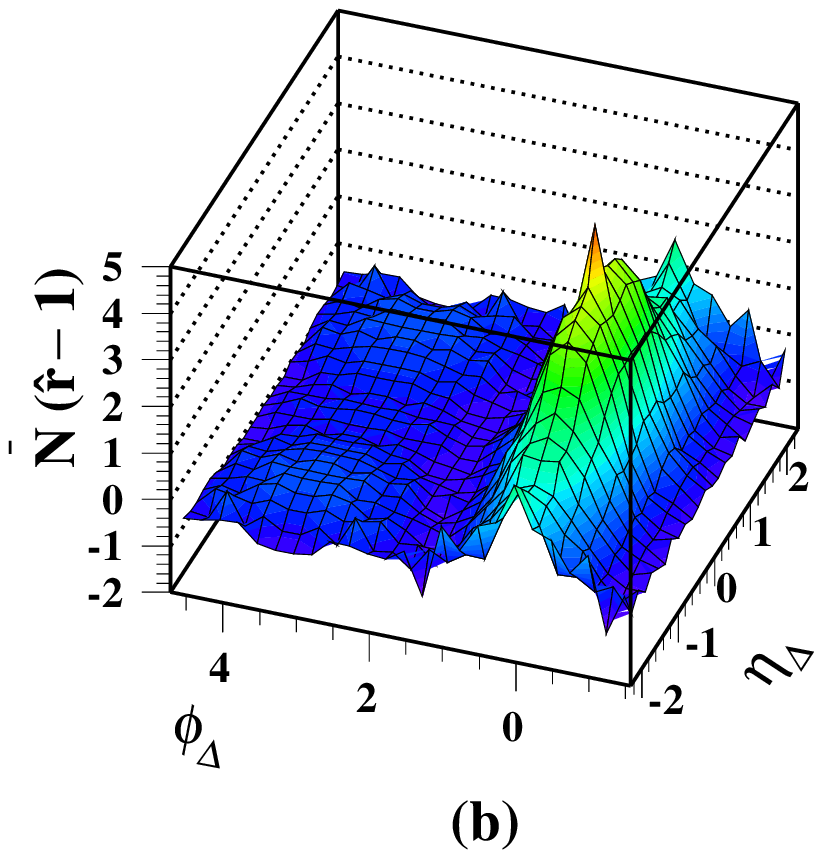}
\includegraphics[keepaspectratio,width=1.65in]{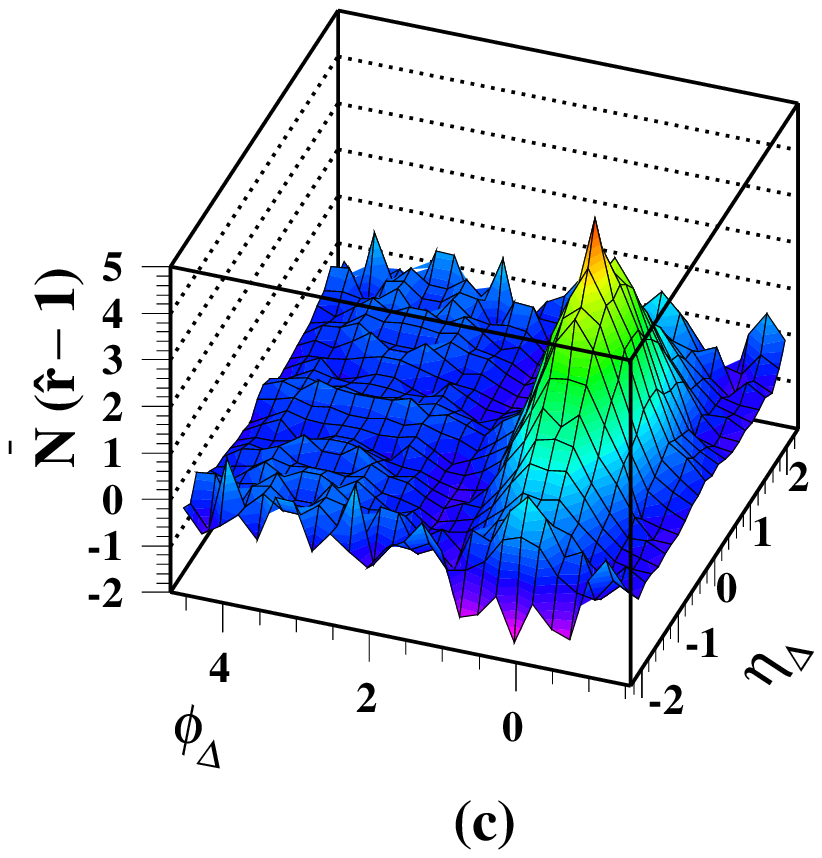}
\includegraphics[keepaspectratio,width=1.65in]{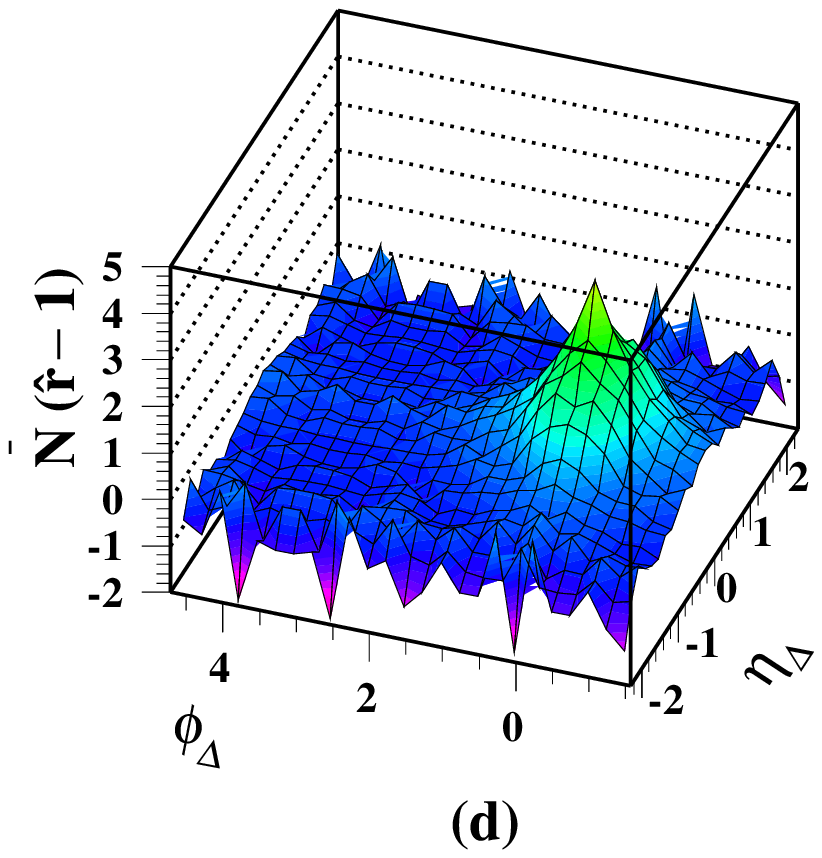}
\caption{\label{Figure2}
The same data as in Fig.~\ref{Figure1}, but with $\eta_\Delta$-independent dipole and quadrupole components on $\phi_\Delta$ (see text) subtracted to reveal `same-side' ($|\phi_\Delta| < \pi / 2$) structures which can be associated with minijets.}
\end{figure}

\section{Two-particle Distributions}

Plotted in Fig.~\ref{Figure1}  are perspective views of CI joint autocorrelations of quantity $\bar N(\hat r - 1)$ (measuring the density of correlated pairs {\em per final-state particle}~\cite{cdpaper}, typically $O(1)$ for all centralities, $\bar N$ is the mean multiplicity in the acceptance) for four centrality classes of Au-Au collisions. $\bar N(\hat r - 1)$ would be independent of centrality if Au-Au collisions were linear superpositions of p-p collisions (participant scaling). The distributions in Fig.~\ref{Figure1} are dominated by 1) a 1D quadrupole component $\propto \cos(2\phi_\Delta)$ conventionally attributed to elliptic flow; 2) a 1D dipole component $\propto \cos(\phi_\Delta)$ associated with transverse momentum conservation in a thermal system, and 3) a 2D `same-side' ($|\phi_\Delta| < \pi / 2$) peak.  The same-side peak is assumed to be associated with parton fragmentation to hadrons, albeit for fragments with much lower $p_t$s than are considered in a conventional jet analysis. Since no leading or trigger particle is invoked in this model-independent analysis we refer to the corresponding fragment systems as {\em minijets} (minimum-bias jets).

Momentum conservation in a thermal multiparticle system should result in a dipole $\cos(\phi_\Delta)$ angular correlation~\cite{fesh}. We also expect back-to-back or {\em away-side} ($\phi_\Delta \sim \pi$) azimuth correlations from momentum conservation in parton scattering (dijets). However, at low $p_t$ the away-side jet structure is broad, and indistinguishable from the dipole $cos(\phi_\Delta)$ component describing momentum conservation in the bulk system. We subtract dipole and quadrupole $cos(2\phi_\Delta)$ components from distributions in Fig.~\ref{Figure1} to obtain Fig.~\ref{Figure2} by minimizing $\eta_\Delta$-independent sinusoidal residuals on the away side region ($|\phi_\Delta| > \pi / 2$) and for $|\eta_\Delta| \sim 2$. The small excess in the (0,0) bins is due to conversion-electron pair contamination. The same-side 2D peaks in this figure are the main subject of this analysis. 

We observe that the away-side region in Fig.~\ref{Figure2} is featureless, even for the most peripheral collisions. If Lund-model strings~\cite{lund} remained dynamically relevant in the final stage of heavy ion collisions we would expect, in the accepted $p_t$ interval, significant correlation structure on the away side of Fig.~\ref{Figure2}: a prominent 1D gaussian on $\eta_\Delta$ approximately symmetric on azimuth and due to local charge conservation on $z$ (during longitudinal string fragmentation) and coupling of $z$ to $\eta$ due to longitudinal Hubble expansion, as observed in p-p collisions~\cite{isrpp,jeffpp}. The absence of such structure suggests that longitudinal strings play {\em no significant role} in the final stage of Au-Au collisions, even for the most peripheral collisions in this study. That trend is consistent with the centrality dependence of net-charge correlations in which structure characteristic of string fragmentation is strongly suppressed with increasing centrality of Au-Au collisions~\cite{cdpaper}. 

\begin{figure}[h]
\includegraphics[width=1.65in,height=1.3in]{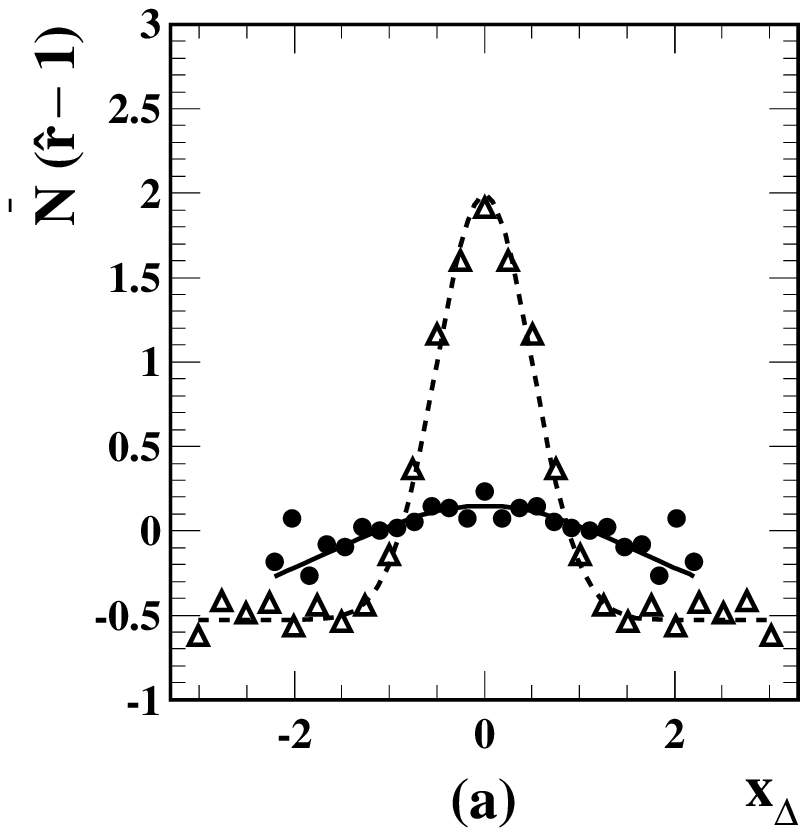}
\includegraphics[width=1.65in,height=1.3in]{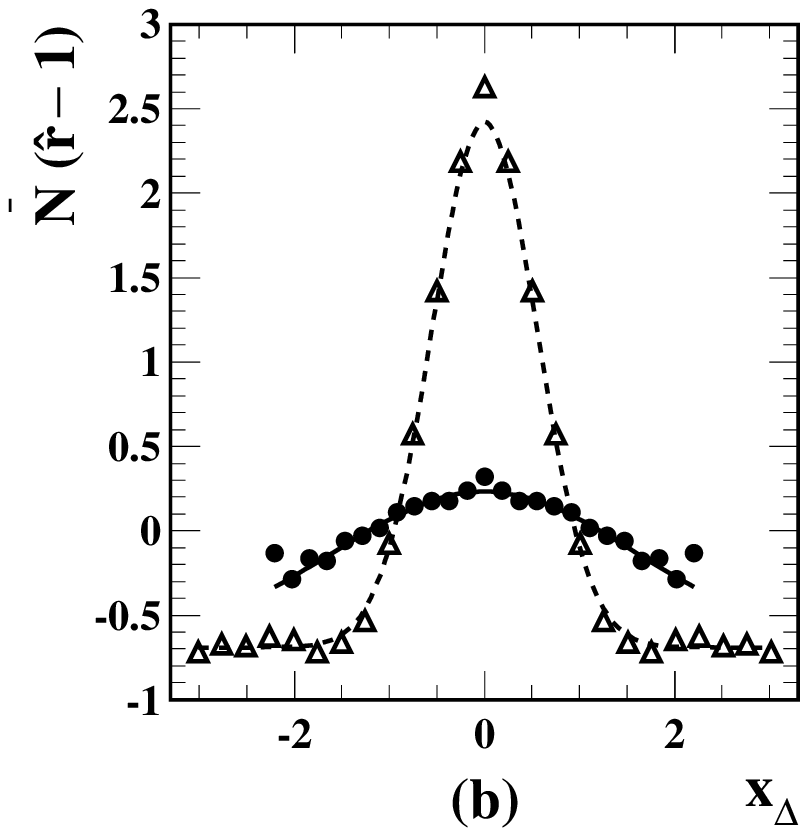}
\includegraphics[width=1.65in,height=1.3in]{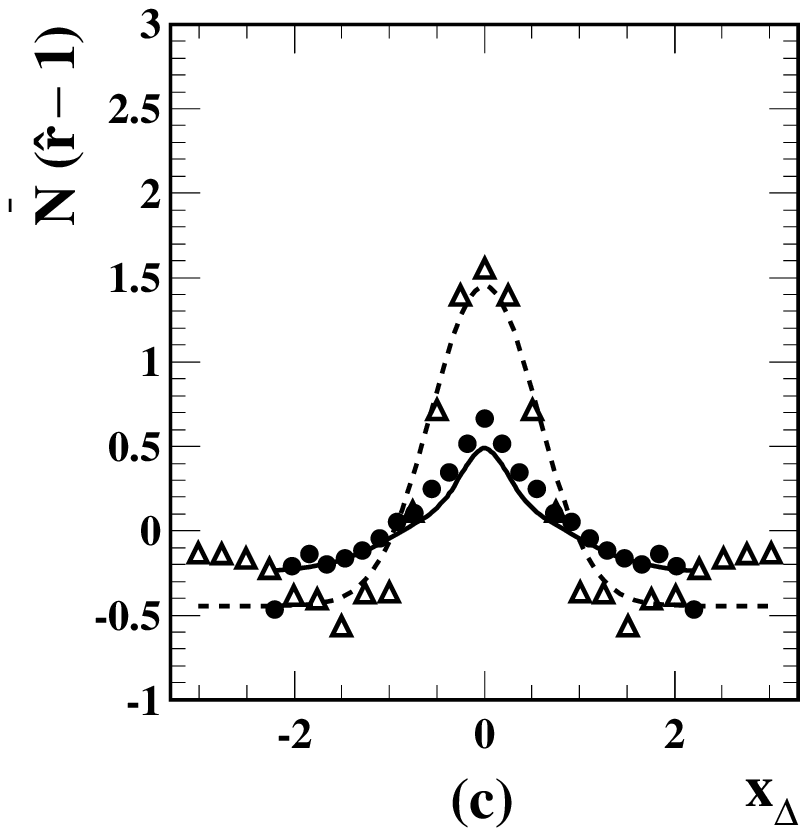}
\includegraphics[width=1.65in,height=1.3in]{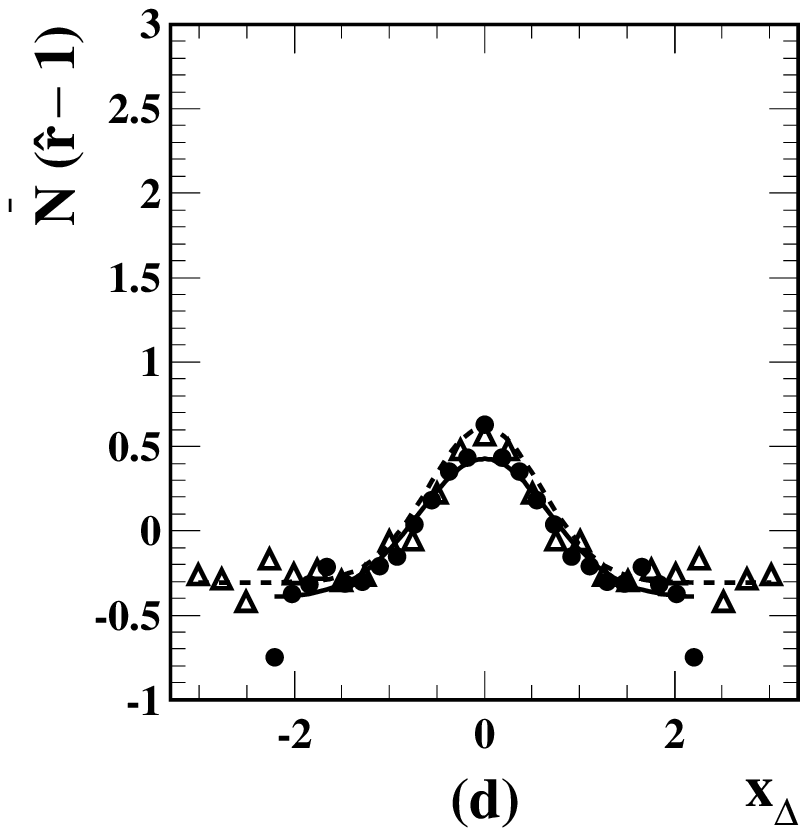}
\caption{\label{Figure3} 
Projections of 2D CI joint autocorrelations $\bar N(\hat{r} - 1 )$
in Fig.~\ref{Figure2} onto difference variables $\eta_{\Delta}$ (solid dots) and $\phi_{\Delta}$ (open triangles). The solid (dashed) curves represent corresponding projections of 2D analytical model fits to the data. 2D peaks are substantially reduced in amplitude by 1D projections.}
\end{figure}

The same-side peak isolated in Au-Au collisions by the multipole subtraction varies strongly with centrality, transitioning from significant elongation on azimuth difference $\phi_\Delta$ for p-p collisions~\cite{jeffpp} to dramatic broadening along $\eta_\Delta$ for the more central Au-Au collisions (note the non-unit aspect ratio of these 2D plots). HBT and Coulomb pair cuts (see Sec.~\ref{SecData}) reduce the bins nearest (0,0) by 10\% or less. 1D projections of data and 2D model fits (discussed below) onto difference variables $\phi_{\Delta}$ and $\eta_{\Delta}$ are shown in Fig.~\ref{Figure3}. Solid dots and curves (open triangles and dashed curves) correspond to $\eta_{\Delta}$ $(\phi_{\Delta})$ projections. 

\section{Errors}

Statistical errors for joint autocorrelations approximately double as $|\eta_{\Delta}|$ increases from 0 to 2 because of the bounded $\eta$ acceptance, but are uniform on $\phi_{\Delta}$ because the azimuthal acceptance is continuous (periodic) in STAR.  Statistical errors for $\hat r$ at $|\eta_{\Delta}| = 0$ vary from $0.0001$ for central collisions to $0.001$ for peripheral collisions. Statistical errors for $\bar N(\hat r - 1)$ ($\sim 0.1$) are nearly independent of centrality. Systematic errors were estimated as in \cite{meanptprl}. Contamination from photon conversions to $e^{\pm}$ pairs is significant only within the bin defined by $|\eta_\Delta| < 0.1$, $|\phi_\Delta| < 0.1$ which was excluded from model fits. The dominant source of systematic error is non-primary background, mainly weak-decay daughters~\cite{spectra}, whose correlation with primary particles is unknown and is estimated by assuming those correlations vary from zero to the measured correlation amplitude for primary particles \cite{meanptprl}. Total systematic errors for data presented in Fig.~\ref{Figure1} are $\pm$7\% of signal, but increase to $\pm$8\% for $|\eta_\Delta| < 0.5$ and to $\pm$11\% for $|\phi_\Delta| < 0.05$. Correlations from resonance $(\rho^0 , \omega )$ decays are $\sim$3\% of the peaks at (0,0) in Fig.~\ref{Figure2} in $|\eta_{\Delta}| < 0.5$, $|\phi_{\Delta}| < 2$~\cite{mevsim}.

\section{Model Fits}

Joint autocorrelations, as in Fig.~\ref{Figure1} but without factor $\bar N$, were fitted with a model function consisting of dipole and quadrupole terms on $\phi_\Delta$, a 1D gaussian on $\eta_{\Delta}$ and a 2D same-side gaussian on $(\eta_{\Delta},\phi_{\Delta})$, plus a constant offset 
\bea \label{Eq3}
F & = & A_{\phi_\Delta}\, \cos(\phi_\Delta) + A_{2\phi_\Delta}\, \cos(2\, \phi_\Delta) + A_0\, e^{- \left( \frac{\eta_{\Delta}}{\sqrt{2} \sigma_{0}} \right)^2 } \nonumber \\ 
&+&  A_1 \, e^{- \left\{ \left( \frac{\phi_{\Delta}}{\sqrt{2} \sigma_{\phi_{\Delta}}} \right)^2  + \left( \frac{\eta_{\Delta}}{\sqrt{2} \sigma_{\eta_{\Delta}}} \right)^2 \right\} } + A_2.
\eea
Best-fit parameters for the model fits shown in Fig.~\ref{Figure3} are listed in Table~\ref{TableI}, including mean multiplicity factor $\bar N$ as in Figs.~\ref{Figure1}-\ref{Figure3} plus tracking efficiency correction factor ${\cal S}$~\cite{norm}. Those fit parameters confirm that with increasing centrality 2D peak structures exhibit 1) strong and non-monotonic amplitude variation, 2) strong $\eta_\Delta$ width increase and 3) significant $\phi_\Delta$ width {\em reduction}. 
\begin{table}[h]
\caption{\label{TableI}
Parameters and fitting errors (only) for model fits [Eq.~(\ref{Eq3})] to
joint autocorrelation data in Fig.~\ref{Figure1} for centrality bins (a) - (d)
(central - peripheral). Total systematic error for efficiency-corrected amplitudes is 11\%~\cite{norm}.}
\begin{tabular}{|c|c|c|c|c|c|} \hline  \vspace{-.13in} & & & & & \\
centrality  & (d) & (c)  & (b)  & (a)  & error\footnote{Range of fitting errors
in percent from peripheral to central.}(\%) \\
\hline \vspace{-.13in} & & & & & \\
${\cal S}$\cite{norm} &   1.19    &   1.22    &  1.25    &   1.27   & 8 (syst)
 \\
$\bar{N}$  &   115.5   &   424.9   &  790.2   &   983.0  &  \\
\hline  \vspace{-.12in} & & & & & \\
${\cal S} \bar{N} A_1$   & 1.93   & 3.23  &  3.72  &  3.10  &   5-2   \\
$\sigma_{\phi_{\Delta}}$ & 0.61  &  0.55  &  0.54  &  0.53  &   4-2   \\
$\sigma_{\eta_{\Delta}}$ & 0.58  &  1.05  &  1.34  &  1.36   &  5-2   \\
\hline \vspace{-.12in} & & & & & \\
${\cal S} \bar{N} A_0$   & 0.60   & 0.32  &  ---  &  ---  &  0.16-0.1\footnote{Magnitude of fitting errors.} \\
$\sigma_{0}$ & 1.11  &  0.24  &  ---  &  ---   &  28-22  \\
\hline \vspace{-.12in} & & & & & \\
${\cal S} \bar{N} A_{2}$     & -0.67 & -0.55  & -0.67  & -0.58 &  0\footnote{Fixed by normalization of $\hat r$}  \\
\hline \vspace{-.12in} & & & & & \\
${\cal S} \bar{N} A_{\phi_\Delta}$   & -0.31 & -0.76  & -0.97  & -0.74 &  22-5   \\
${\cal S} \bar{N} A_{2\phi_\Delta}$   & 1.05  &  2.72  &  1.30  &  0.32 &  2-17   \\
\hline  \vspace{-.12in} & & & & & \\
$\chi^2$/DoF & $\frac{439}{316}$ & $\frac{419}{316}$ & $\frac{675}{316}$ & $\frac{415}{316}$ &  \\
\hline
\end{tabular}
\end{table}

\section{Discussion}

\begin{figure}[h]
\includegraphics[width=1.6in,height=1.5in]{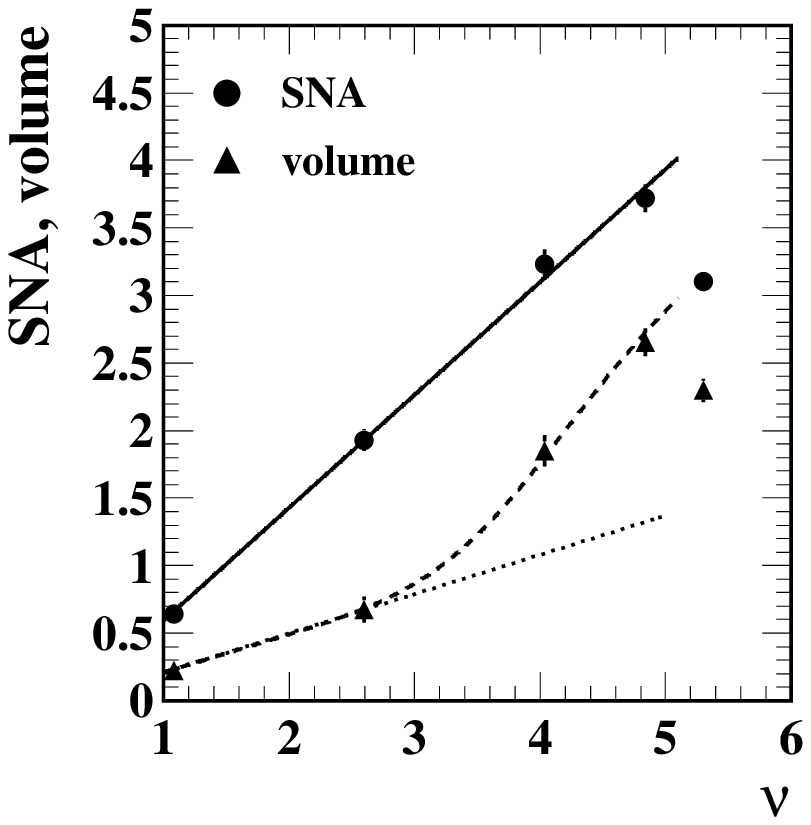}
\includegraphics[width=1.6in,height=1.5in]{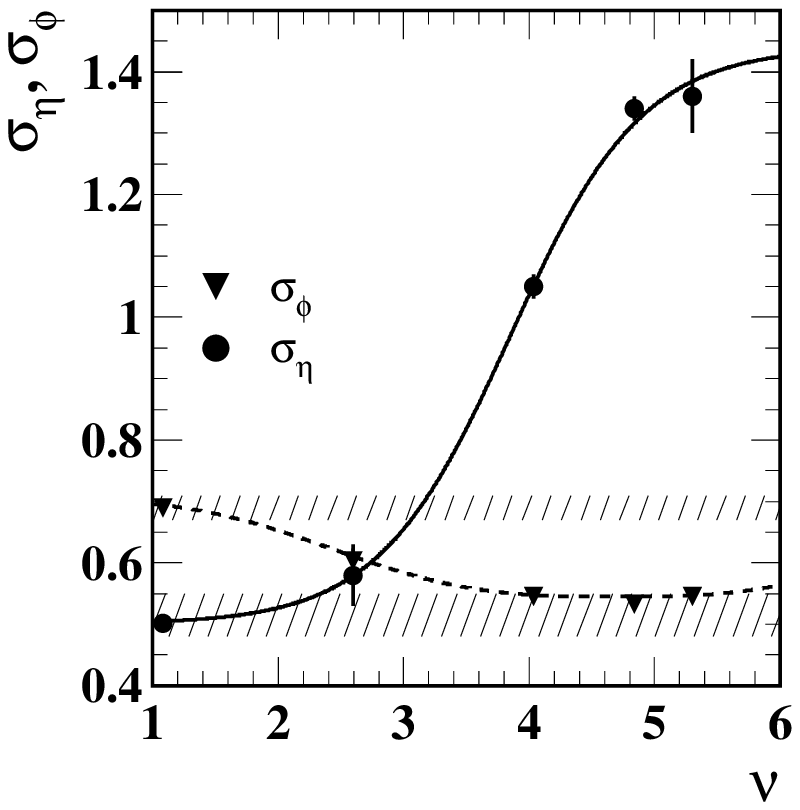}
\caption{\label{Figure4}
Left: Fitted amplitudes and volumes for peaks in Fig.~\ref{Figure2} plotted {\em vs} mean participant path length $\nu$~\cite{nu}, from Table~\ref{TableI}. Right: Fitted widths $\sigma_{\eta_\Delta}$ (dots) and $ \sigma_{\phi_\Delta}$ (triangles).  Hatched regions show p-p values.  Curves guide the eye.}
\end{figure}

In Fig.~\ref{Figure4}, the same-side peak amplitudes and widths from model fits are plotted $vs$ centrality measure $\nu$~\cite{nu} (mean participant path length in terms of the number of encountered nucleons, $\approx 2 N_{bin} / N_{part}$) along with measurements obtained from p-p collision data~\cite{jeffpp}. The same-side peaks in Fig.~\ref{Figure2} differ strongly from those for p-p collisions, where for the latter a 2D gaussian peak significantly elongated on {\em azimuth} dominates the same-side structure, widths on $\eta_\Delta$ and $\phi_\Delta$ being $\sim$ 0.5 and 0.7 respectively~\cite{jeffpp}. The similar same-side peak widths for mid-peripheral Au-Au collisions in this analysis (panel~(d), $\nu \sim 2.5$) are consistent with the p-p result. In central Au-Au collisions however, the widths of the same-side peak reverse the sense of the asymmetry: the peak is dramatically elongated on $\eta_\Delta$, the width ratio increasing to 2.6. 

Same-side peak amplitudes ${\cal S}\bar N A_1$ (measuring correlations per final-state particle) in the left panel of Fig.~\ref{Figure4} increase nearly linearly with path-length as expected for {\em independent} binary collisions. However, peak {\em volume} $\equiv  {\cal S}\bar N A_1\, \sigma_{\eta_\Delta}\, \sigma_{\phi_\Delta}$ ($\propto$ minijet fragment number) has a more complex variation, strongly departing from linear $\nu$ scaling (dotted line) above $\nu = 2.5$. The dashed curve in the left panel is derived from the curves describing amplitude and peak widths. The volume excess beyond the linear extrapolation may indicate the onset of a strongly dissipative medium in which more correlated fragments with less $p_t$ result from each scattered parton. The latter increase is very likely a lower-$p_t$ manifestation of the observed suppression of the high-$p_t$ part of the $p_t$ spectrum measured by quantity $R_{AA}$~\cite{raa}. It is notable that the peak {\em amplitude} does not deviate from a linear trend, except for the most central point.

We speculate that the mechanism modifying the same-side peak in central Au-Au collisions is strong coupling of {\em minimum-bias} energetic partons (no high-$p_t$ trigger is imposed) to a {\em longitudinally-expanding} colored medium developed in the more central Au-Au collisions. Hadrons from fragmenting partons sample the {\em local velocity structure} of the pre-hadronic parton-medium coupled system. Growth of the colored medium and its coupling to the fragmenting parton with increasing collision centrality is then indicated by increased width on $\eta_\Delta$ of the same-side correlation peak.


The perturbative QCD expectation for angular correlations about the jet  thrust axis in p-p collisions corresponds to a {\em nearly-symmetric} near-side peak on $(\eta_\Delta,\phi_\Delta)$. That `{\em in vacuo}\,' result is indeed observed in p-p collisions for higher-$p_t$ fragments ($> 2.5$ GeV/c). However, in a minimum-bias autocorrelation analysis of p-p data~\cite{jeffpp} {\em strong deviations} from expected pQCD angular symmetry about the jet thrust axis are observed. The Hijing Monte Carlo collision model~\cite{jetquench} includes a conventional pQCD model of jet production and quenching in A-A collisions. The default Hijing same-side peak is observed to be symmetric, and the widths on $\eta_\Delta$ {\em and} $\phi_\Delta$ both increase by only 10\% when jet quenching is imposed~\cite{ptscale}, {\em seriously underpredicting} the large width increase on $\eta_\Delta$ and contradicting the width {\em decrease} on $\phi_\Delta$ observed in the present analysis of Au-Au data. The pQCD jet-quenching mechanism in Hijing cannot produce an asymmetry on $(\eta_\Delta,\phi_\Delta)$, given the symmetry about the parton momentum of its perturbative bremsstrahlung quenching mechanism. Prominent low-$p_t$ longitudinal string-fragment correlations on $\eta_\Delta$ are observed for all Hijing centralities,  also contradicting results of the present analysis noted above.  RQMD~\cite{rqmd} CI correlations are featureless except for small flow-related correlations on $\phi_\Delta$.

Recently, effects of a flowing medium on parton energy loss and fragmentation have been explored theoretically~\cite{salgado}. The premise of that study is that gluon bremsstrahlung from energetic partons transiting a colored medium should be sensitive to the local structure of the velocity field on the medium. The model considered is uniform medium flow (`directed flow') transverse to the energetic parton momentum. A static medium is expected to broaden the bremsstrahlung angular distribution and hence the near-side peak (symmetric broadening from jet quenching is observed already with Hijing~\cite{ptscale}). Medium flow transverse to the parton direction was found in~\cite{salgado} to shift and distort the fragment-energy angular distribution relative to the thrust axis. In the LHC context, for 100 GeV jets with typical {\em energy} angular width $\sim 0.05$, the effect of the flowing medium on the angular distribution was found to be comparable to the width magnitudes (`marked medium-induced deviations'). However, the absolute angular changes were small.

In the RHIC context a comparison was made with a STAR leading-particle analysis of jet correlations~\cite{fuqiang}. The prediction of~\cite{salgado} for trigger particles with $p_t \in$ [4,6] GeV/c is width variation from peripheral to central 200 GeV Au-Au collisions of 0.35 (symmetric) to 0.4 on azimuth and to 0.56 on pseudorapidity. Those width increases are similar in magnitude to the symmetric Hijing width increases noted above. However, they differ qualitatively from the width {\em decrease} from 0.7 to 0.5 on azimuth and the dramatic width increase from 0.5 to 1.4 on pseudorapidity observed in the present minimum-bias jet study. The study of directed (vector) flow and parton bremsstrahlung in~\cite{salgado} does not address the issue of longitudinal Hubble (Bjorken) {\em expansion}, a tensor aspect of the velocity field. Coupling of parton fragmentation to the velocity field may be much stronger than what can be modelled perturbatively, requiring a nonperturbative treatment. The analysis in~\cite{salgado} also does not address the centrality dependence of angular deformation, which is strongly nonlinear on path length as demonstrated in Fig.~\ref{Figure4} (right panel).

Fluctuation analysis has been generally advocated as a probe of heavy ion collisions. It is important to note that charge-independent number fluctuations observed within a given detector acceptance integrate {\em over that acceptance} the CI joint autocorrelations presented in this paper (within a constant offset), as described in~\cite{inverse}. The constant offset in the joint autocorrelations, dominated by participant (`volume') fluctuations, is easily separated from the differential angular structure which reveals details of nuclear collision dynamics.  


\section{Summary}

In conclusion, we have for the first time measured charge-independent joint autocorrelations on difference variables $\phi_{\Delta}$ {\em and} $\eta_{\Delta}$ for Au+Au collisions at $\sqrt{s_{NN}}$ = 130 GeV. Low-$p_t$ string-fragment correlations which appear prominently in p-p collisions are not observed for any centrality in this study: longitudinal string degrees of freedom are apparently strongly suppressed even for fairly peripheral Au-Au collisions. Other correlation structures are observed to have substantial amplitudes. In addition to azimuth structures associated with elliptic flow and transverse momentum conservation we observe a near-side peak structure varying from a nearly-symmetric shape on ($\eta_\Delta,\phi_\Delta$) in peripheral collisions to a shape strongly elongated on $\eta_\Delta$ in central collisions. We interpret the same-side peak as resulting from fragmentation of minimum-bias partons observed with no trigger condition (minijets). The trend of minijet angular deformation, observed in this first jet analysis with {\em low}-$p_t$ hadrons, can be interpreted as a transition from {\em in vacuo} jet fragmentation in p-p and peripheral Au-Au collisions to strong coupling of minimum-bias partons to a longitudinally-expanding colored medium in the more central collisions as part of a parton dissipation process. The concept of parton energy loss in heavy ion collisions is thereby extended to strongly {\em nonperturbative} aspects. Detailed comparisons between data and pQCD-based theory reveal several qualitative differences.

We thank the RHIC Operations Group and RCF at BNL, and the
NERSC Center at LBNL for their support. This work was supported
in part by the HENP Divisions of the Office of Science of the U.S.
DOE; the U.S. NSF; the BMBF of Germany; IN2P3, RA, RPL, and
EMN of France; EPSRC of the United Kingdom; FAPESP of Brazil;
the Russian Ministry of Science and Technology; the Ministry of
Education and the NNSFC of China; IRP and GA of the Czech Republic,
FOM of the Netherlands, DAE, DST, and CSIR of the Government
of India; Swiss NSF; the Polish State Committee for Scientific 
Research; STAA of Slovakia, and the Korea Sci. \& Eng. Foundation.


\begin{thebibliography}{99}
\bibitem{trackcuts}
Track cuts required minimum track separations 5, 12 and 20~cm at radii 50, 127 and 200~cm from the TPC axis.  Crossing pairs with separations less than 10~cm ($z$) and 30~cm (azimuth) at mid-radius were also excluded.
\bibitem{endpoint}
$N_0$, the half-max point at the end of the minimum-bias
distribution plotted as $d\sigma / dN_{ch}^{1/4}$, is an estimator on multiplicity $N$ for the maximum number of participant nucleons; $N/N_0 \simeq N_{part}/N_{part,max}$ within 4\%.
\bibitem{nu} $\nu$ estimates the mean participant path length as a number of encountered nucleons. For this analysis $\nu \equiv 5.5 \, (N/N_0)^{1/3} \simeq 5.5\, (N_{part}/N_{part,max})^{1/3} \simeq 2N_{bin}/N_{part}$, based on Glauber-model simulations. $N_{part}$ is the number of participants and $N_{bin}$ is the number of binary collisions.
\bibitem{norm} Extrapolation factors ${\cal S}$ for $\bar{N}(\hat{r} - 1)$~\cite{meanptprl} correct for contamination and inefficiency~\cite{spectra}. Systematic error in ${\cal S}$ was estimated to be $\pm$8\%.  Total systematic error for extrapolated quantities in Table~\ref{TableI} was 11\%.
\end{thebibliography}
\end{document}